\documentclass[aps,prb,reprint]{revtex4-2}
\usepackage{graphicx}
\usepackage{amsmath}
\usepackage{amssymb}
\begin{document}
\title{Ferroelectricity in strained Hf$_2$CF$_2$ monolayer}
\author{Ziwen Wang}
\author{Ning Ding}
\author{Churen Gui}
\author{Shanshan Wang}
\author{Ming An}
\email{Email: amorn@seu.edu.cn}
\author{Shuai Dong}
\email{Email: sdong@seu.edu.cn}
\affiliation{School of Physics, Southeast University, Nanjing 211189, China}
\date{\today}

\begin{abstract}
Low dimensional ferroelectrics are highly desired for applications and full of exotic physics. Here a functionalized MXene Hf$_2$CF$_2$ monolayer is theoretically studied, which manifests a nonpolar to polar transition upon moderate biaxial compressive strain. Accompanying this structural transition, a metal-semiconductor transition occurs. The in-plane shift of unilateral fluorine layer leads to a polarization pointing out-of-plane. Such ferroelectricity is unconventional, similar to the recently-proposed interlayer-sliding ferroelectricity but not identical. Due to its specific hexapetalous potential energy profile, the possible ferroelectric switching paths and domain walls are nontrivial, which are mediated via the metallic paraelectric state. In this sense, the metallic walls can be manipulated by reshaping the ferroelectric domains.
\end{abstract}
\maketitle

\section{Introduction}
In ferroelectric materials, charge dipoles form spontaneous orders and generate macroscopic polarizations, which can be switched by external fields. Therefore, ferroelectric materials have high technical values for applications, including nonvolatile memories, sensors, as well as photonics devices \cite{Scott2007,Rabe2007}. In recent years, ferroelectrics have also been discovered in two-dimensional (2D) materials \cite{Chang274,Belianinov2015}. Their atomic-scale thickness and passivated surfaces perfectly match the demands of device miniaturization \cite{Shuang2021}. In addition, unconventional ferroelectric physics beyond that in three-dimensional (3D) crystals have also been revealed \cite{An2020,Guan2020,184104}. Along this promising route, it is essential to discover more 2D ferroelectrics with exotic physical mechanisms and better performances.

MXene is a new branch of the 2D materials tree \cite{Naguib2011,Khazaei2013}, which can be synthesized from its precursor MAX phase generally known as layered ternary carbides and nitrides \cite{Barsoum2000}. Since the first report in 2011, more than 30 MXenes have been experimentally obtained \cite{Gogotsi2019}. Due to their diverse compositions and impressive versatility, MXenes have attracted many research attentions \cite{Gogotsi2019,Hantanasirisakul2018}. The ferroelectricity in MXenes has also been investigated. For example, Zhang \textit{et al.} predicted that Hf$_2$VC$_2$F$_2$ is the first 2D type-II multiferroic material, whose polarization is generated by noncollinear spin texture \cite{Zhang2018}. Also, the oxygen-functionalized scandium carbide MXene (Sc$_2$CO$_2$) was predicted to be a 2D ferroelectric with both in-plane and out-of-plane polarization \cite{Chandrasekaran2017}. Its ferroelectricity is mainly driven by the functionalized oxygen ions, and plenty physics like antiferroelectricity as well as metallic domain boundary was predicted \cite{Chandrasekaran2017}.

In addition to functionalization, mechanical strain is another effective way to modulate materials' properties. In fact, the strain-induced or strain-modulated ferroelectricity has been extensively studied in 3D crystals \cite{Zeches977,Lee2010,Zhao2014}. For 2D materials, the strain effects have been used to modulate magnetism, band structure, and electronic effective mass \cite{Pereira2009,Peelaers2012,Zhao2014}. However, the study of strain-induced ferroelectrocity in 2D materials remains rare.

In this work, based on density functional theory (DFT) calculations, the effect of biaxial strain on a fluorine-functionalized MXene Hf$_2$CF$_2$ has been systematically investigated. Our calculations find a first-order nonpolar-polar structural phase transition (also metal-semiconductor transition) in Hf$_2$CF$_2$ monolayer under moderate compressive strain. Driven by the in-plane shifting of unilateral fluorine layer, an out-of-plane ferroelectric polarization is generated. Comparing with its sister compound Sc$_2$CO$_2$ \cite{Chandrasekaran2017}, its ferroelectric performance is better, with a larger polarization and a lower energy barrier for switching. Furthermore, the understanding of underlying ferroelectric physics is deepened, going beyond the previous claim. The in-plane polarization is unambiguously ruled out in this family, and its nontrivial domain structure is clarified.

\section{Computational methods}
Our DFT calculations were performed using Vienna \textit{ab initio} Simulation Package (VASP) \cite{Kresse1996}. The electron-ion interactions were described by projector-augmented-wave (PAW) pseudopotentials with semi-core states treated as valence states \cite{Bl1994}. Plane-wave cutoff energy was fixed as $500$ eV. The exchange and correlation were treated using Perdew-Burke-Ernzerhof (PBE) parametrization of the generalized gradient approximation (GGA). A $\Gamma$-centered $13\times13\times1$ Monkhorst-Pack \textit{k}-mesh was adopted for Brillouin zone sampling. Since the GGA usually underestimates band gaps, the more precise Heyd-Scuseria-Ernzerhof (HSE06) functional was also adopted for comparison \cite{Heyd2006}.

To simulate a monolayer, a vacuum layer more than $20$ \AA{} was added along the $c$-axis direction to avoid the interaction between two neighboring slices. Both the in-plane lattice constant and atomic positions were fully optimized iteratively until the Hellmann-Feynman force on each atom and the total energy were converged to $10^{-3}$ eV/\AA{} and $10^{-7}$ eV, respectively.

To simulate the biaxial strain effects, two approaches are adopted. The simpler one is to directly shrink the in-plane lattice constant. The more realistic one is to use MoS$_2$ monolayer or van der Waals (vdW) bulk as the substrate. Then the atomic positions were further optimized accordingly. The vdW interatomic forces are described by the D2 Grimme method \cite{Grimme2006}.

The ferroelectric polarization was calculated using the Berry phase method \cite{King-Smith1993}. The possible switching paths between different states were evaluated using the nudged elastic band (NEB) method \cite{Henkelman2000}.

\begin{figure}
\includegraphics[width=0.48\textwidth]{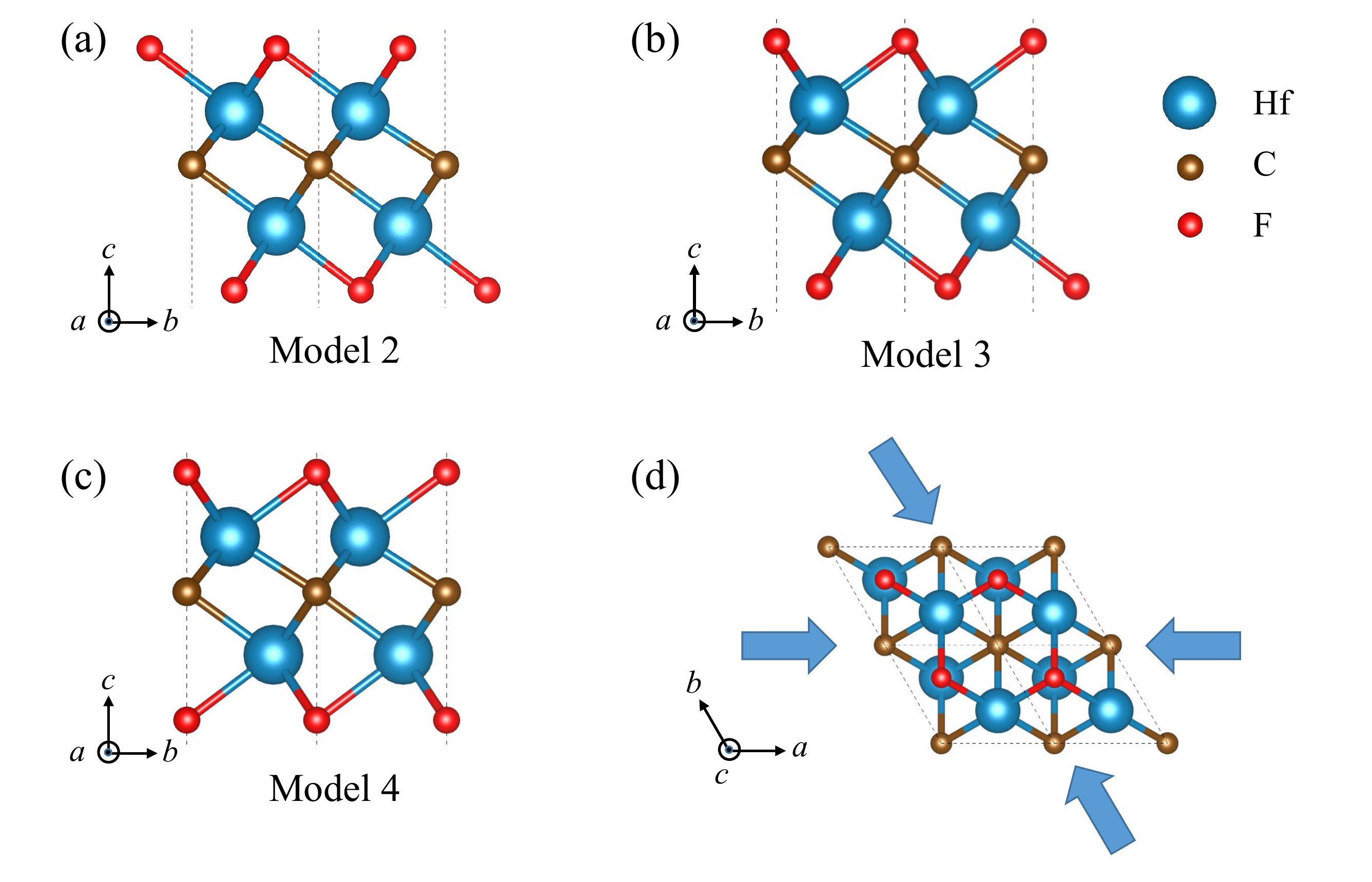}
\caption{Possible structures of MXene Hf$_2$CF$_2$. (a-c) Side views of different configurations of Hf$_2$CF$_2$. (d) Top view of the model 2. The biaxial compressive strain is indicated by arrows.}
\label{F1}
\end{figure}

\section{Results and discussion}
\subsection{Structural transition driven by strain}
In general, MXenes are exfoliated from the precursor MAX phases by selective chemical etching. During the exfoliation process, the surfaces of MXenes are usually passivated by fluorine, oxygen, or other chemical ligands \cite{Naguib2011,Anasori2015}. Here, Hf$_2$C monolayer can be obtained from its parent bulk Hf$_2$AlC, which has been synthesized in experiment \cite{Lapauw2016}. For fully fluorinated Hf$_2$C monolayer, i.e., Hf$_2$CF$_2$, there are four most possible configurations for chemical terminations, as proposed in Ref.~\cite{Khazaei2013}.

According to our structural optimization, the model 1, as shown in Fig. S1 in Supplemental Materials (SM) \cite{SM}, is dynamic unstable, which will spontaneously transform to the model 2. Therefore, only the models 2, 3, and 4 will be discussed in the following. The differences among these three configurations can be clearly distinguished from the side view, as shown in Fig.~\ref{F1}(a-c), which mainly lie in the positions of surface fluorine ions relative to the central carbon. The top view of model 2 is also illustrated in Fig.~\ref{F1}(d).

Basic physical properties of Hf$_2$CF$_2$ monolayer obtained from DFT calculations are summarized in Table~\ref{Tab-1}. For comparison, the parent Hf$_2$AlC bulk is also calculated, whose lattice constants match the experimental values ($a$=$3.271$ \& $c$=$14.363$ \AA{} \cite{Lapauw2016}) very well. Also our DFT results of Hf$_2$CF$_2$ monolayer are highly consistent with previous theoretical values (e.g., $a=3.264$, $3.232$, and $3.179$ \AA{} for the models 2, 3, and 4 in Ref.~\cite{ur2016}).

For the free standing Hf$_2$CF$_2$ monolayer, the model 2 owns the lowest energy among all considered configurations, while the model 3 is slightly higher in energy. Both models 2 and 4 are metallic, while the model 3 is a semiconductor. All these results agree well with previous first-principles studies \cite{Khazaei2013,ur2016}, which indicates the reliability of our calculations. It is noted that the space group of model 3 is a sub-group of model 2, by breaking the inversion symmetry along $c$-axis. Thus, the model 3 is polar, while the model 2 is nonpolar.

\begin{table}
\caption{Basic physical properties of Hf$_2$CF$_2$ and its parent phase Hf$_2$AlC, obtained in our DFT calculations. The lattice constants $a$ and $c$ are in unit of \AA. The energy are in relative to the model 2 and in unit of meV/f.u.. The band gap is in unit of eV. Only the model 3 is polar and semiconducting.}
\begin{tabular*}{0.48\textwidth}{@{\extracolsep{\fill}}lccccc}
\hline \hline
 & space group & $a$ & $c$ & energy & band gap                              \\
\hline
Hf$_2$AlC & $P6_3/mmc$ (no. 194)   & $3.273$  & $14.379$  & -      & -      \\
Model 2   & $P\bar{3}m1$ (no. 164) &  $3.263$ & -         & $0$    & 0      \\
Model 3   & $P3m1$ (no. 156)       & $3.232$  & -         & $23.5$ & $0.31$ \\
Model 4   & $P\bar{3}m1$ (no. 164) & $3.181$  & -         & $304$  & 0      \\
\hline \hline
\end{tabular*}
\label{Tab-1}
\end{table}

\begin{figure}
\includegraphics[width=0.48\textwidth]{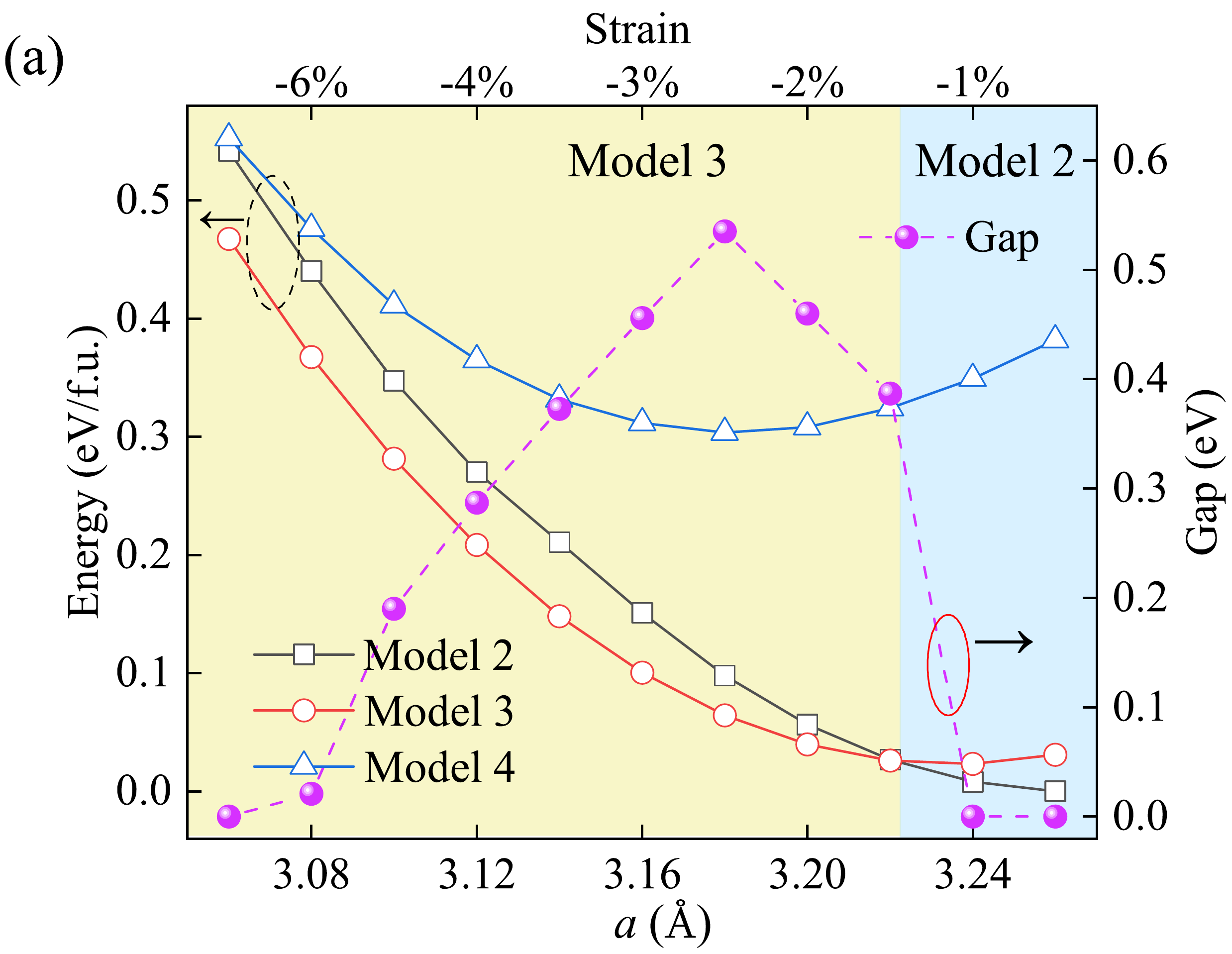}
\includegraphics[width=0.48\textwidth]{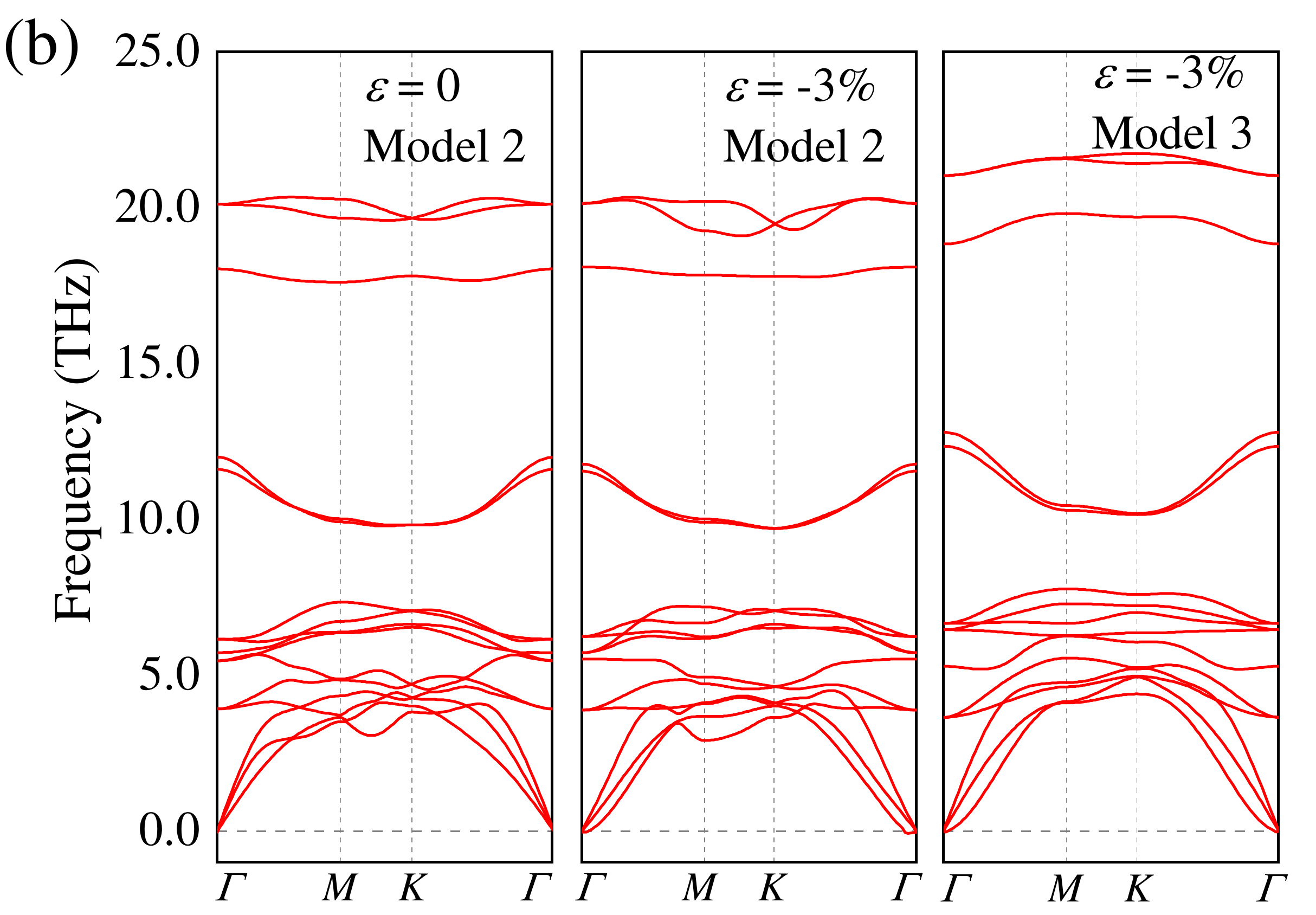}
\caption{(a) Ground state phase diagram of Hf$_2$CF$_2$ as a function of lattice constant. Left axis: energy per f.u.. The energy of optimized model 2 is taken as the reference. Right axis: band gap. Upper axis: the corresponding biaxial strain, defined as $\varepsilon=(a-a_0)/a_0$, where $a_0$ and $a$ are the in-plane lattice constant of the pristine (model 2) and strained MXene, respectively. (b) Phonon spectra of the models 2 and 3, under different biaxial compressive strains. The model 2 is dynamically stable without strain (left) and under $\varepsilon=-3\%$ compressive strain (middle). The model 3 is also stable in this condition (right).}
\label{F2}
\end{figure}

The proximate energies of the models 2 and 3 provide the possibility to pursuit the switching functionality of their contrastive physical properties (e.g., nonpolar {\it vs} polar, metallic {\it vs} semiconducting). Figure~\ref{F2}(a) shows the evolutions of models 2-4 as a function of in-plane lattice constant $a$, i.e., to simulate the biaxial strain. Obviously, the compressive strain is more advantageous for the models 3 and 4, since their optimized in-plane lattice constants are smaller than that of model 2. Indeed, the energy of model 3 becomes lower than that of model 2 when the biaxial strain is over $-1.3\%$. In contrast, the energy of model 4 is always higher than other two configurations within the considered range of biaxial strain. The phase transition between the models 2 and 3 is accompanied by a metal-semiconductor transition, with a sudden change of band gap from $0$ to $0.39$ eV. With further  compression, the band gap of model 3 increases first and then decreases to $0$ gradually, as shown in Fig.~\ref{F2}(a).

Besides the factor of energy, the dynamic stability is also essential for structural transition. To determine the stability of Hf$_2$CF$_2$ monolayer, the phonon spectra are calculated, as shown in Fig.~\ref{F2}(b). No imaginary phonon frequency emerges in all three cases: unstrained model 2, strained models 2 and 3 (at $\varepsilon=-3\%$), implying their dynamic stability. In other words, the model 2 remains (meta)-stable in the strained condition, which is an evidence of the first-order transition.

\begin{figure}
\includegraphics[width=0.48\textwidth]{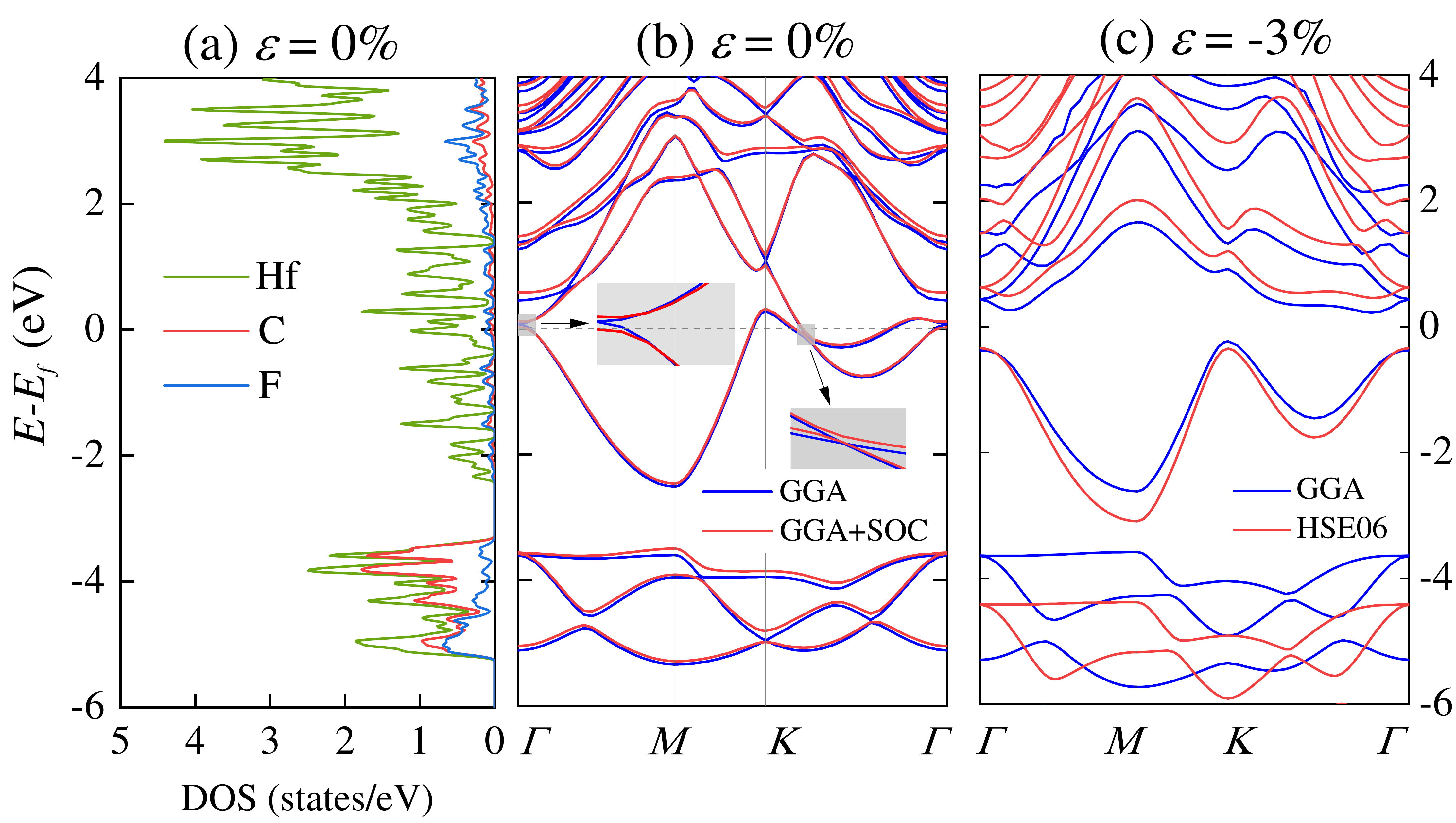}
\caption{The representative electronic structures of Hf$_2$CF$_2$, which show a metal-semiconductor transition upon compressive strain. (a-b) Unstrained model 2. (a) DOS shows that the electronic state around the Fermi level is mostly from Hf's $5d$ orbitals. (b) The corresponding band structure. Blue: without spin-orbit coupling (SOC); Red: with SOC. (c) Band structure for strained model 3, in which an indirect band gap is opened. The HSE06 result is similar to the GGA one, with a slightly larger band gap.}
\label{F3}
\end{figure}

To explore the strain effect on electronic properties of Hf$_2$CF$_2$ monolayer, the density of states (DOS) and band structures under different compressive strains are shown in Fig.~\ref{F3}. Initially, the model 2 exhibits metallic behavior, and the bands around the Fermi level is mostly contributed by Hf's $5d$ orbitals [Fig.~\ref{F3}(a)]. Since the spin-orbit coupling (SOC) of $5d$ orbitals is prominent, the band structure is recalculated with SOC enabled, presented in Fig.~\ref{F3}(b) for comparison. However, the SOC effect is almost negligible and the system remains metallic. Under a moderate biaxial strain, the model 3 becomes the ground state, which opens an indirect band gap, as shown in Fig.~\ref{F3}(c). For comparison, the HSE06 band structure is also shown in Fig.~\ref{F3}(c), which is similar to the GGA one with a slightly enlarged band gap. The HSE06 band structure for unstrained model 2 can be found in Fig. S2 of SM \cite{SM}, which does not alter its electronic structure too much. Thus, in the following, only the GGA calculations are presented.

For transition metal elements, partially filled $d$ orbitals may lead to magnetism. However, our DFT calculations do not find any local magnetic moment on Hf$^{3+}$ ion, even if a moderate $U$ is applied. The absence of local magnetic moment is due to the large bandwidth of Hf's $5d$ orbitals (see Fig.~\ref{F3} for example).

In practice, strain is usually achieved by proper substrates. Here the MoS$_2$ is chosen as the substrate to mimic the strain effect in real heterostructures, because of their approximate lattice constant ($3.181$ \AA{} for MoS$_2$, and $3.263$ \AA{} for Hf$_2$CF$_2$) and similar structure (hexagonal in-plane symmetry). The slight smaller lattice constant of MoS$_2$ can provide a compressive strain to the attached Hf$_2$CF$_2$ monolayer, as requested to stabilize the model 3. Our DFT calculation shows that the model 3 configuration indeed has a lower energy than model 2 for Hf$_2$CF$_2$ monolayer on MoS$_2$ substrate. More details regarding the MoS$_2$ substrate can be found in SM (Fig. S3 and Table S1) \cite{SM}.

\subsection{Ferroelectricity}
\begin{figure}
\includegraphics[width=0.48\textwidth]{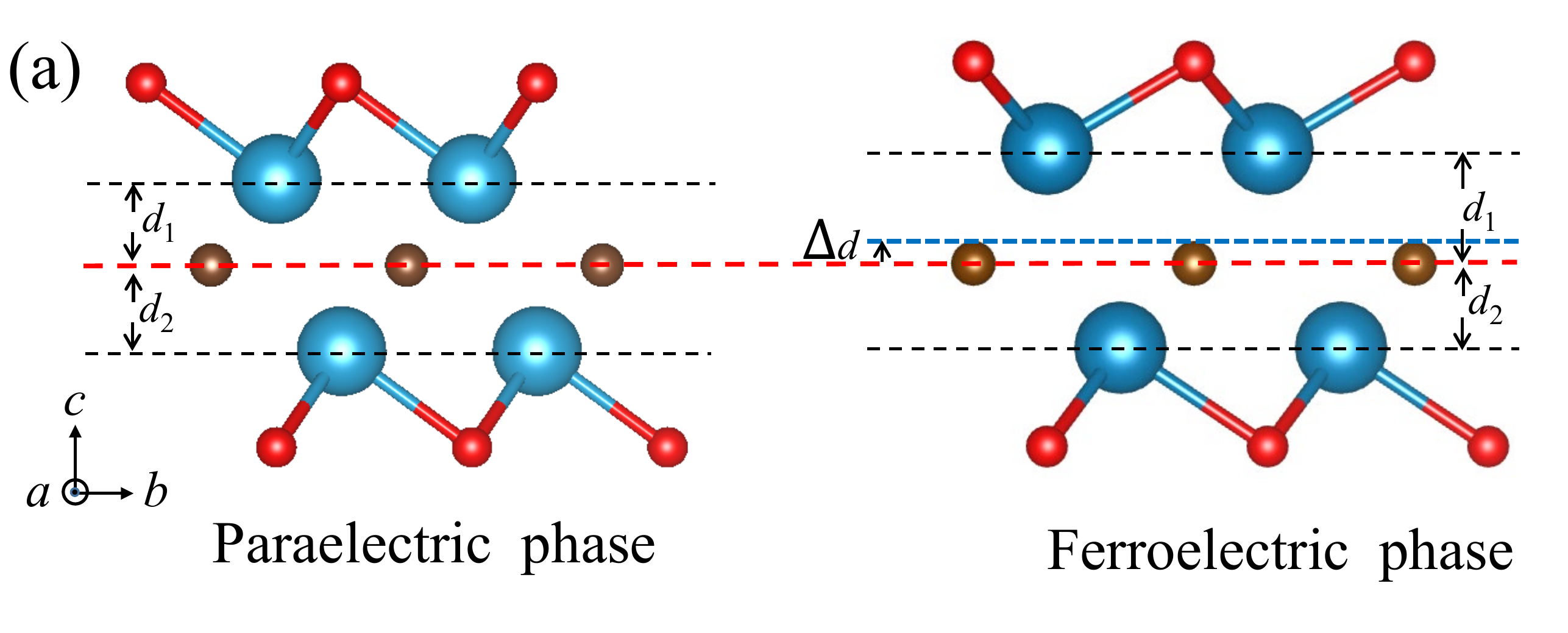}
\includegraphics[width=0.48\textwidth]{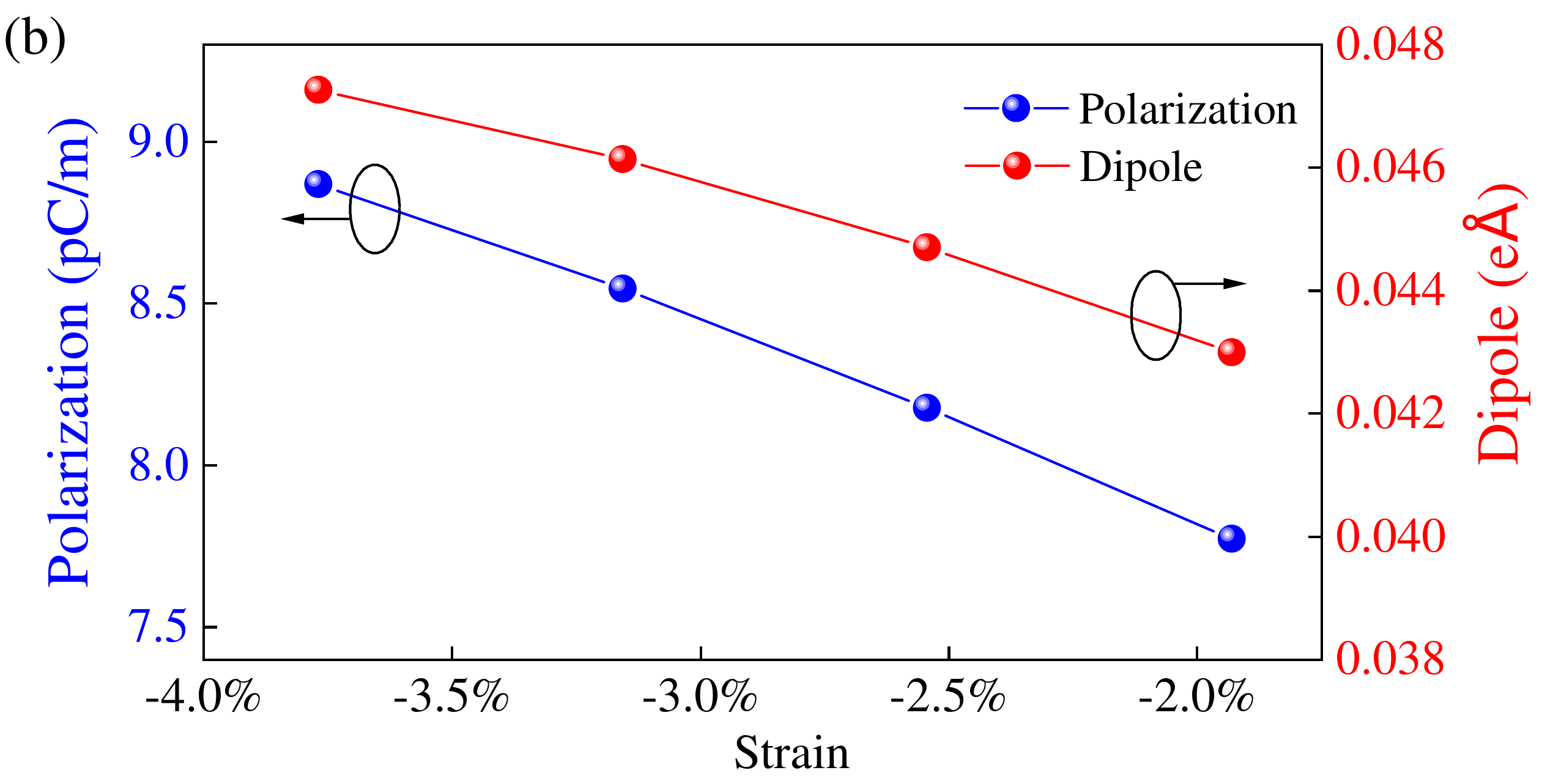}
\includegraphics[width=0.48\textwidth]{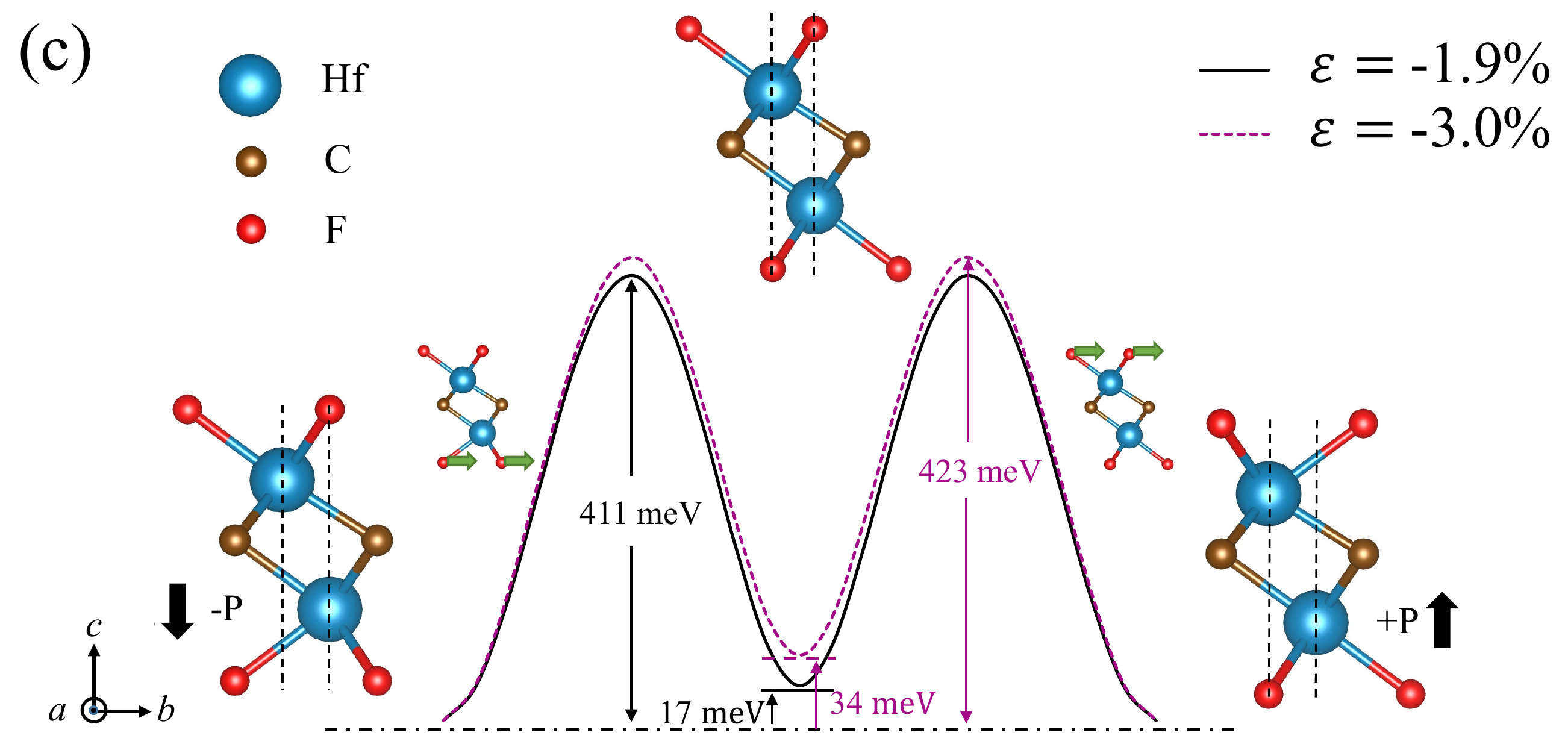}
\caption{(a) Comparison between the paraelectric phase and ferroelectric phase. The C layer is aligned as the common reference level. The Hf-C layer distances are symmetric in the paraelectric phase (i.e., $d_1=d_2$), which become asymmetric in the ferroelectric one (i.e., $d_1\neq d_2$). Blue broken line: the middle of Hf bilayer. (b) The 2D polarization and associated dipole moment as a function of compressive strain. (c) The energy barrier of possible ferroelectric switching path for Hf$_2$CF$_2$ at compressive conditions (solid curve: $\varepsilon=-1.9\%$ and broken curve: $\varepsilon=-3\%$).}
\label{F4}
\end{figure}

According to the symmetry analysis, the structure of model 3 is polar but model 2 is nonpolar. For the model 3, the upper Hf is just above the lower F, while the lower Hf is not below the upper F. In contrast, the F and Hf columns are always synchronized in the model 2. Such in-plane shift of unilateral F layer breaks the inversion symmetry along the $c$-axis: the distances between Hf and C layers become different for the upper ($d_1$) and lower ($d_2$) ones. For example, at $\varepsilon$=$-1.9\%$, $d_1$=$1.333$ \AA{} and $d_2$=$1.156$ \AA{}, while the original $d_1$=$d_2$=$1.219$ \AA, as visualized in Fig.~\ref{F4}(a). Then an out-of-plane polarization is induced. The atomic positions at $\varepsilon$=$-1.9\%$ for the models 2 and 3 are summarized in Table~\ref{Tab-2}.

\begin{table}
\caption{The atomic positions of Hf$_2$CF$_2$ in the models 2 and 3 at a compressive strain $\varepsilon$=$-1.9\%$. Noting that the lattice constant along $c$-axis is fixed as $30$ \AA{} (including the vacuum layer). The $c$-axis position of C is fixed at the middle of the cell.}
\begin{tabular*}{0.48\textwidth}{@{\extracolsep{\fill}}lcc}
 \hline \hline
 &Wyckoff positions& Internal coordinates\\
 \hline
 Model 2     &  & \\
 \hline
 C     & $1b$            &($0$, $0$, $1/2$)\\
 F     & $2d$            &($2/3$, $1/3$, $0.4139$)\\
 Hf    & $2d$            &($1/3$, $2/3$, $0.4594$) \\
 \hline
Model 3      & & \\
\hline
 C      & $1a$     &($0$, $0$, $1/2$)        \\
 F (1)  & $1a$     &($0$, $0$, $0.5904$)       \\
 F (2)  & $1c$     &($2/3$, $1/3$, $0.4161$)   \\
 Hf (1) & $1c$     &($2/3$, $1/3$, $0.5444$)     \\
 Hf (2) & $1b$     &($1/3$, $2/3$, $0.4614$)     \\
\hline \hline
\end{tabular*}
\label{Tab-2}
\end{table}

In this sense, such origin of ferroelectricity in the model 3 is similar to the so-called interlayer-sliding ferroelectricity \cite{Li2017}, although here the sliding occurs at the surface layers, not between the vdW layers in their proposal. Due to the stronger interactions of F-Hf-C bonding, here a moderate polarization $7.8$ $p$C/m (equals to $1.48$ $\mu$C/cm$^2$ in the 3D case by considering its thickness) is obtained at $\varepsilon$=$-1.9\%$, larger than the vdW interlayer-sliding ferroelectric BN ($\sim0.68$ $\mu$C/cm$^2$ \cite{Li2017}) and thus easier to be detected in future experiments. This ferroelectric polarization can be slightly enhanced with increasing compressive strain, as shown in Fig.~\ref{F4}(b). This tendency is also reasonable: stronger compressive strain, looser structure along the $c$-axis, which is advantageous for ferroelectric displacements.

The ferroelectric switching process is also simulated using the NEB method, as shown in Fig.~\ref{F4}(c). Since the paraelectric state is dynamically stable, the energy profile is a triple well one, instead of conventional double well. At $\varepsilon$=$-1.9\%$, the energy barrier between the ferroelectric state and paraelectric state is relative higher $\sim410$ meV/f.u., although the energy difference between these two phases is only $17$ meV/f.u.. Noting that this value should be considered as a theoretical upper limit, since the real switching process probably finds an easier path than our NEB one. Even though, this value is already lower than that of Sc$_2$CO$_2$ ($\sim520$ meV \cite{Chandrasekaran2017}), and comparable to the most extensively studied multiferroic BiFeO$_3$ ($427$ meV) \cite{Theoretical2006}. Furthermore, with increasing compressive strain, the middle well becomes slightly shallower. The ferroelectric properties, including the polarization and switching process, are not affected by the MoS$_2$ substrate \cite{SM}.

Since the paraelectric state (as the intermediate state of ferroelectric switching path) is metallic, it is necessary to clarify the feasibility of switching by electric field. Usually, the ferroelectric switching needs a full insulating path. However, for some improper ferroelectric systems, metallicity can be allowed to coexist with polarity \cite{Zhang2015}. In addition, the concept of ``2D hyperferroelectric metals" was also proposed, with a polarization pointing out-of-plane \cite{Luo2017}. Furthermore, Fei \textit{et al.} experimentally realized the ferroelectric switching of a 2D metal WTe$_2$ using gate electrodes \cite{Fei2018}. Thus, for 2D ferroelectrics, the switching path can be compatible with the metallicity, at least in the nanoscale limit.

Finally, it should be noted that the in-plane shifting of F layers will not induce any in-plane polarization, which is beyond the intuitional expectation and different from previous claim \cite{Chandrasekaran2017}. Both the paraelectric model 2 and ferroelectric model 3 have the $C_3$ rotation symmetry, which forbids any in-plane polarization. In fact, the displacement of F ion starts from a high-symmetric point to another high-symmetric point, which can create a polarization quantum and will not induce a real polarization \cite{SPALDIN20122}.

\subsection{Ferroelectric domain structure}

\begin{figure}
\includegraphics[width=0.48\textwidth]{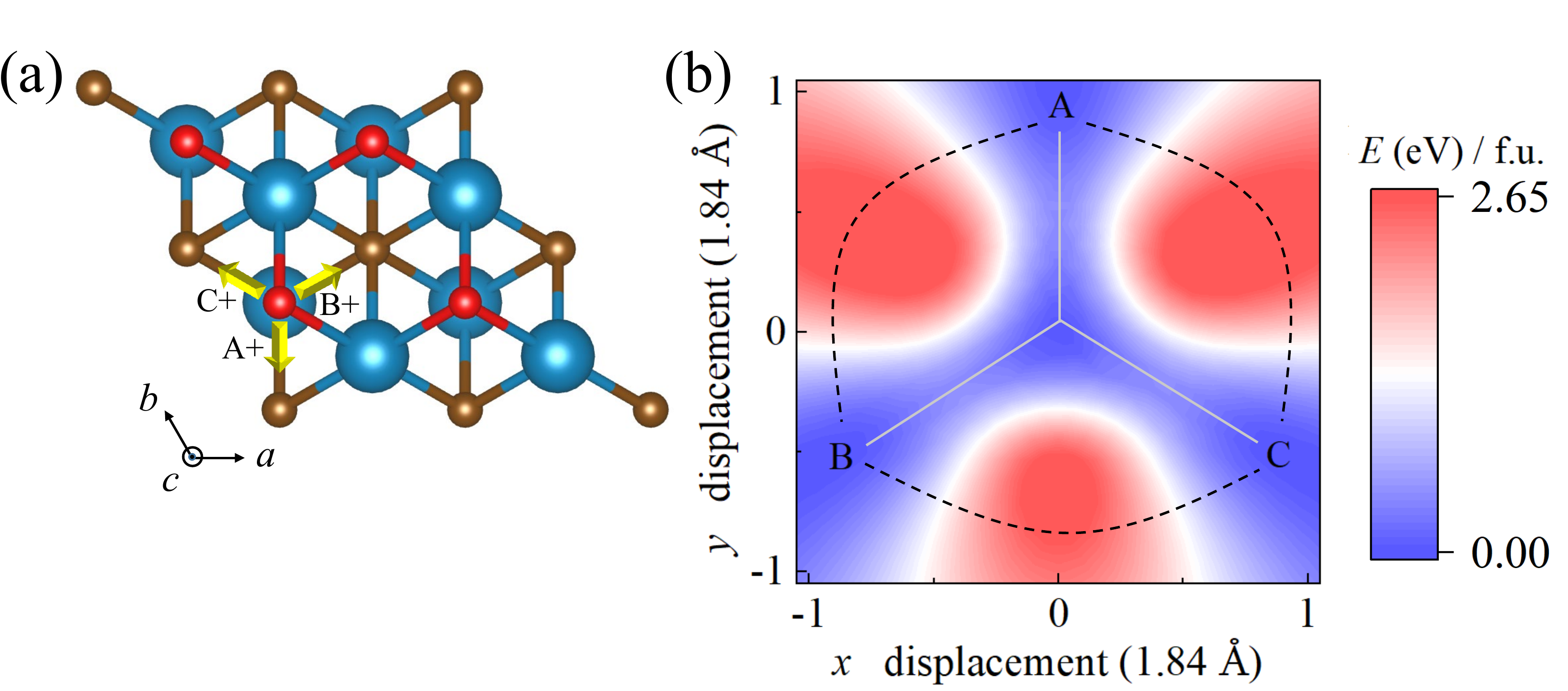}
\includegraphics[width=0.48\textwidth]{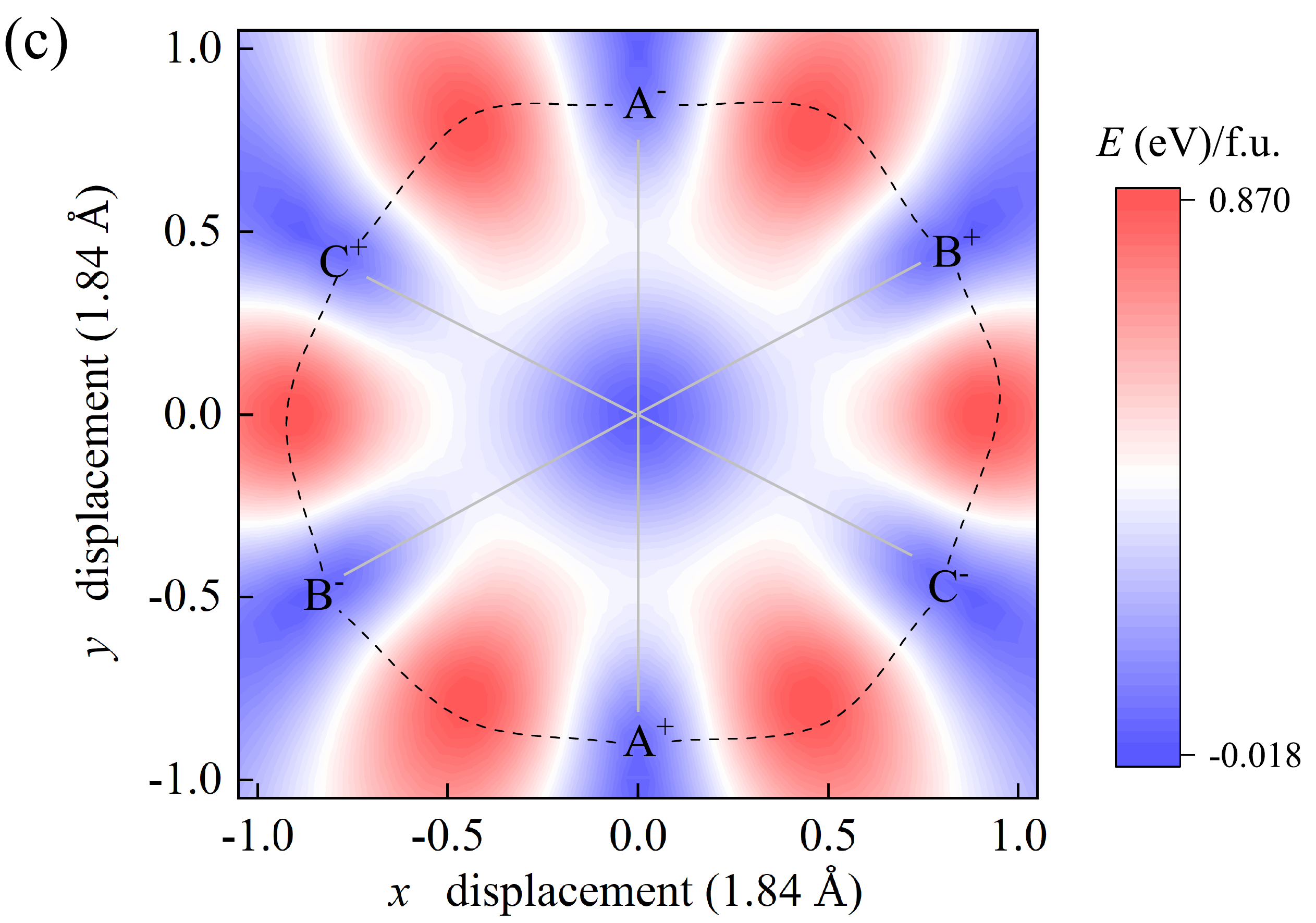}
\caption{(a) For each F layer, there are three equivalent shifting directions (named as A, B, C and indicated by yellow arrows) from the paraelectric phase to the ferroelectric phase. (b-c) The energy landscapes as a function of F displacements, obtained using the NBE method. The +/- indicates the shifting for upper/lower F layer, which are also the sign of polarization (up/down). Here two plots are presented: (b) only involves a single F-layer shifting and (c) contains two F-layers shifting. The A/B/C domains are always isolated by high energy barriers.}
\label{F5}
\end{figure}

Considering the $C_3$ (i.e., the in-plane triple rotation) symmetry of paraelectric phase, there are three possible shifting directions for each F plane, denoted as A/B/C, as shown in Fig.~\ref{F5}(a). These triple directions plus two F layers provide the $3\times2$ degrees of freedom, similar to the hexagonal improper ferroelectric $R$MnO$_3$ and $R$FeO$_3$ \cite{Choi2010}. In those hexagonal oxides, the triple choices of Mn or Fe trimerizations (denoted as $\alpha$, $\beta$, $\gamma$) and ferroelectric polarization ($+$ or $-$) generate six possible domains, which form the $\mathbb{Z}_3\times\mathbb{Z}_2$ topological antiphase domain walls (i.e. the $\alpha^+$-$\beta^-$-$\gamma^+$-$\alpha^-$-$\beta^+$-$\gamma^-$ vortices). The origin of such topological domain vortices in those hexagonal oxides is their Mexican-hat like energy landscape \cite{Artyukhin2014}.

\begin{figure}
\includegraphics[width=0.46\textwidth]{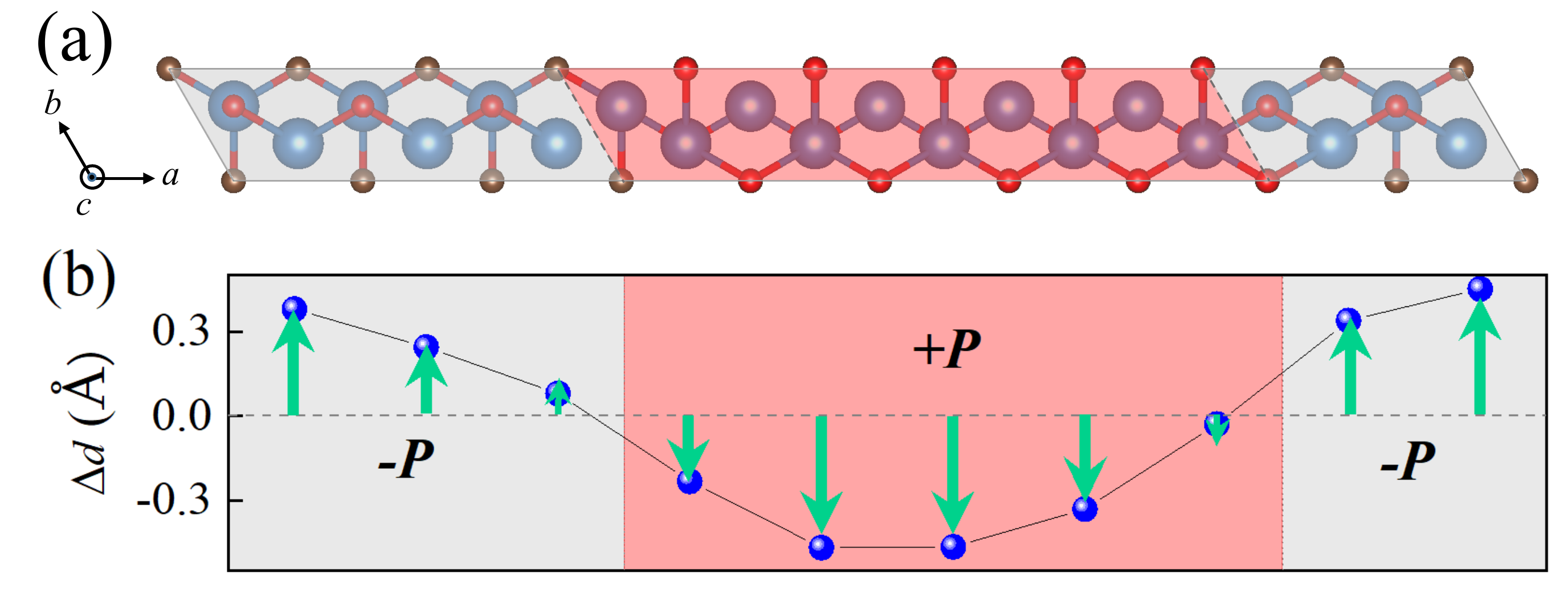}
\includegraphics[width=0.44\textwidth]{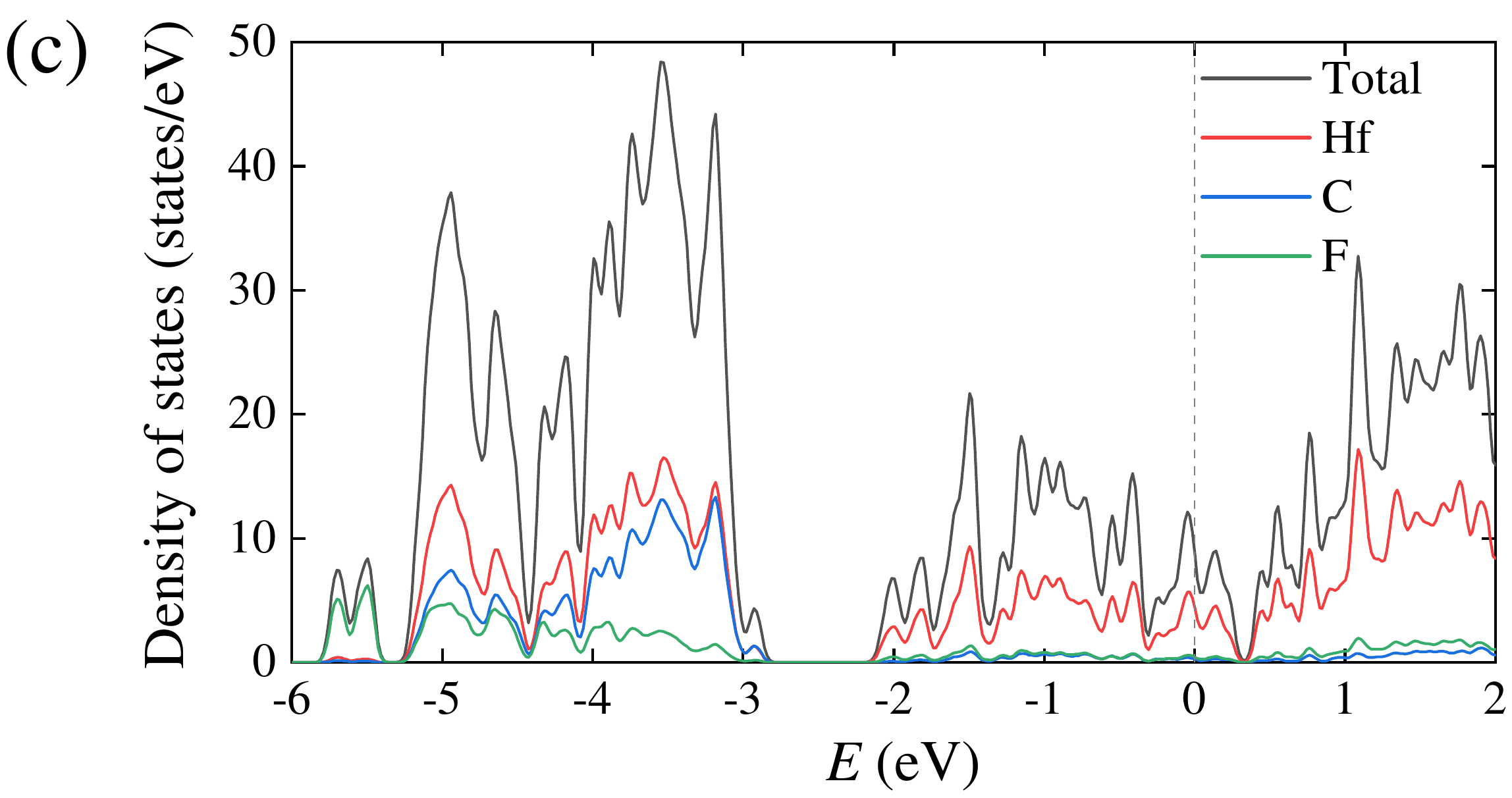}
\includegraphics[width=0.46\textwidth]{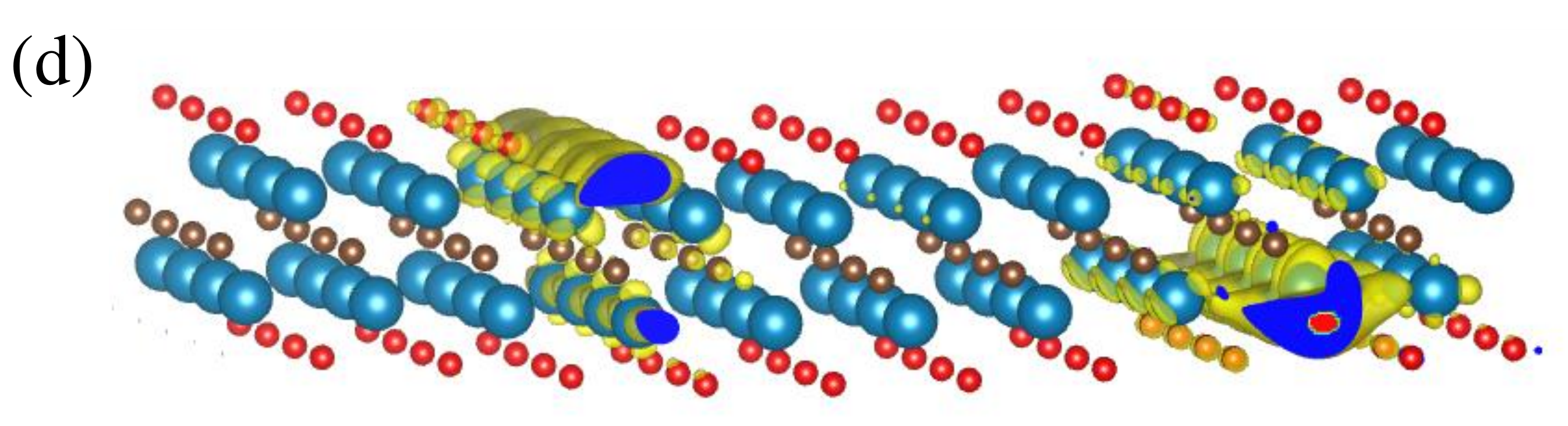}
\caption{(a) Sketch diagram of a ferroelectric domain wall between the A$^-$ and B$^+$ domains. (b) The relaxed domain wall. The local dipole moments, characterized by $\Delta d$=$(d_1-d_2)/2$, are indicated by arrows. The value of $\Delta d$ tends to be zero around the domain wall. (c) The DOS of supercell containing the domain walls between A$^-$ and $B^+$. The metallicity is expected from the domain wall. (d) The corresponding electron density distribution, integrated from $-0.4$ eV to the Fermi level. It is obvious that these electrons only appear at the domain walls.}
\label{F6}
\end{figure}

Then it is interesting to ask whether similar topological domain structure can appear in Hf$_2$CF$_2$ monolayer? To preliminarily answer this question, here the energy landscape as a function of F layer shifting vectors are calculated using the NEB method, as shown in Fig.~\ref{F5}(b-c). Unexpectedly, this energy landscape is totally different from the Mexican-hat one \cite{Artyukhin2014}. The energy barriers of direct paths between adjacent energy minima, i.e., those broken lines in Fig.~\ref{F5}(b-c), are rather high. Then all preferred domain walls will be between the ferroelectric model 3 and paraelectric model 2, i.e., those solid lines in in Fig.~\ref{F5}(b-c), instead of walls between antiphase domains of the model 3. Thus, no topological domain structure is expected here.

To further investigate the domain wall profiles, a ferroelectric domain wall between the A$^-$ and B$^+$ domains is studied, as shown in Fig.~\ref{F6}(a). Starting from a sharp domain boundary, the relaxed structure shows a continuous transition between polarization up and down, as shown in Fig.~\ref{F6}(b). Such continuous modulation of local dipoles is a result of the energy landscape [Fig.~\ref{F5}(c)], different from the very sharp domain wall predicted in WO$_2$Cl$_2$ which owns the quadruple energy wells \cite{Lin2019}. Since the domain wall is close to the model 2 state, it is expected to be metallic, as shown in Fig.~\ref{F6}(c). The electronic state around the Fermi level is visualized in Fig.~\ref{F6}(d), which further confirms the metallicity of domain wall.

The metallicity of ferroelectric domain wall in Hf$_2$CF$_2$ is intrinsic and free from defects. For comparison, the previously studied conductive domain wall in BiFeO$_3$ \cite{Seidel2009,Rojac2017} is mostly due to the accumulation of charged defects and remains semiconducting (but with smaller local band gaps). Thus, the underlying physics involved in BiFeO$_3$ domain wall and Hf$_2$CF$_2$ domain wall is conceptually different. Such metallic wall provides the possibility to pursuit domain wall nanoelectronics \cite{Catalan2012}.

\section{Conclusion}
In summary, the ferroelectricity of strained fluorine-functionalized MXene Hf$_2$CF$_2$ has been systematically studied by DFT calculations. Under a moderate compressive strain, the Hf$_2$CF$_2$ monolayer will transform from the nonpolar $P\bar{3}m1$ phase to polar $P3m1$ phase, accompanied by the metal-semiconductor transition. The in-plane shift of unilateral F ions breaks the inversion symmetry along the $c$-axis, thus generates a vertical polarization. Possible ferroelectric switching paths and domain wall configurations have also been depicted, based on its specific hexapetalous potential energy profile. Its domain wall is not sharp and is intrinsically metallic, which provides a unique opportunity for domain wall nanoelectronics. More following theoretical and experimental works are encouraged to verify our predictions and explore more unconventional ferroelectricity in 2D materials.

\begin{acknowledgments}
We thank J. Chen and S. X. Song for illuminating discussions. This work was supported by National Natural Science Foundation of China (Grant No. 11834002). Most calculations were done on Tianhe-2 at National Supercomputer Centre in Guangzhou and the Big Data Computing Center of Southeast University.
\end{acknowledgments}

\bibliography{reference}
\bibliographystyle{apsrev4-2}
\end{document}